\documentclass[superscriptaddress,reprint]{revtex4-1}  

\RequirePackage{lineno}

\usepackage{amsmath}    
\usepackage{graphicx}   
\usepackage{dcolumn}    
\usepackage{bm}         
\usepackage{lineno}     
\usepackage{hyperref}   
\usepackage{xspace}     
\usepackage{pdfpages}

\begin{document}
\title{Measurement of the Proton-Air Cross Section with Telescope
  Array's Middle Drum Detector and Surface Array in Hybrid Mode.}

\author{R.U.~Abbasi}
\affiliation{High Energy Astrophysics Institute and Department of
Physics and Astronomy, University of Utah, Salt Lake City, Utah, USA}

\author{M.~Abe}
\affiliation{The Graduate School of Science and Engineering, Saitama
University, Saitama, Saitama, Japan}

\author{T.~Abu-Zayyad}  
\affiliation{High Energy Astrophysics Institute and Department of
Physics and Astronomy, University of Utah, Salt Lake City, Utah, USA}

\author{M.~Allen} 
\affiliation{High Energy Astrophysics Institute and Department of
Physics and Astronomy, University of Utah, Salt Lake City, Utah, USA}

\author{R.~Azuma} 
\affiliation{Graduate School of Science and Engineering, Tokyo
Institute of Technology, Meguro, Tokyo, Japan}

\author{E.~Barcikowski} 
\affiliation{High Energy Astrophysics Institute and Department of
Physics and Astronomy, University of Utah, Salt Lake City, Utah, USA}

\author{J.W.~Belz}
\affiliation{High Energy Astrophysics Institute and Department of
Physics and Astronomy, University of Utah, Salt Lake City, Utah, USA}

\author{D.R.~Bergman} 
\affiliation{High Energy Astrophysics Institute and Department of
Physics and Astronomy, University of Utah, Salt Lake City, Utah, USA}

\author{S.A.~Blake} 
\affiliation{High Energy Astrophysics Institute and Department of
Physics and Astronomy, University of Utah, Salt Lake City, Utah, USA}

\author{R.~Cady} 
\affiliation{High Energy Astrophysics Institute and Department of
Physics and Astronomy, University of Utah, Salt Lake City, Utah, USA}

% ...


\begin{abstract}
In this work we are reporting on the measurement of the proton-air
inelastic cross section $\sigma^{\rm inel}_{\rm p-air}$ using the Telescope Array (TA)
 detector. Based on the measurement of the
 $\sigma^{\rm inel}_{\rm p-air}$ the proton-proton cross section
 $\sigma_{\rm p-p}$ value
is also determined  at $\sqrt{s} = 95_{-8}^{+5}$ TeV. Detecting cosmic ray events at ultra high energies
 with Telescope Array 
enables us to study this fundamental parameter that we are otherwise
 unable to access with particle accelerators. The data used in this
 report is the hybrid events observed by 
the Middle Drum fluorescence detector together with the surface array 
detector collected over five years. The value of the  $\sigma^{\rm inel}_{\rm p-air}$ is  found to be equal to  
$ 567.0 \pm 70.5 [{\rm Stat.}]  ^{+29}_{-25} [{\rm Sys.}]$ mb. The total proton-proton cross section 
is subsequently inferred from Glauber Formalism  and Block, Halzen and 
Stanev  QCD inspired fit and is found to be equal to   $170
_{-44}^{+48} [{\rm Stat.}] _{-17}^{+19} [{\rm Sys.}] $mb. 
\end{abstract}
\maketitle

\section{Introduction}

Measuring the proton-air inelastic cross section $\sigma^{\rm inel}_{\rm p-air}$ from cosmic rays at ultra high energies allows us to achieve knowledge of a fundamental particle property that we are unable to attain with measurements at current accelerators. At the time of writing of this paper the highest proton-proton center of mass energy that could be attained by the modern accelerators is $\sim 14$~TeV by the Large Hadron Collider (LHC). However, Ultra High Energy Cosmic Ray (UHECR) experiments have been reporting on the proton-air inelastic cross section starting with the Fly's Eye in 1984 at $\sqrt s = 30$~TeV~\citep{FE1987} and ending with the most recent result of the Auger experiment at $\sqrt s = 57$~TeV in 2012~\citep{Auger2012}.

The current high energy models agree in their predictions of rising
proton-air cross section with energy. The high energy models are in
reasonable agreement at lower energies, below $10^{15}$ eV, where they
are tuned to measurements of multi-particle production provided by
particle accelerators. However the high energy models diverge in
describing fundamental parameters such as hadronic cross sections,
elasticity, and secondary particle multiplicity above 1~PeV where the
models rely solely on theoretical expectations~\citep{Ulrich2010}. Studying the
energy dependence of the proton-air cross section is important in
constraining the extrapolation of the hadronic models to high energy.

Detecting UHECR showers provides the opportunity to study fundamental particle properties.
Optimally, to measure the $\sigma^{\rm inel}_{\rm p-air}$ directly, we observe the first point of the proton-air interaction slant depth $X_{1}$ and fit the distribution of  $X_{1}$ to recover the interaction length $\lambda_{\rm p-air}$.  However, since the observation of the first point of interaction to obtain the nucleon-air cross section is not feasible, the inelastic proton-air cross section is calculated using the distribution of the observed shower maximum $X_{max}$.   $\lambda_{\rm p-air}$ and consequently $\sigma^{\rm inel}_{\rm p-air}$, are derived from the $X_{max}$ distribution's exponential tail. 

In this work, we report on the measurement of the proton-air inelastic
cross section $\sigma^{\rm inel}_{\rm p-air}$ using the Telescope
  Array's Middle Drum detector together with the Surface array
  Detector (MD-SD) in hybrid
  mode data~\citep{Abbasi2014}. The method
  used in this calculation is ``the $K$-Factor method'', were the
  underlying  assumption is a proportionality between the tail of the
  $X_{max}$ distribution and $X_{1}$. Details of the method, the
  result, and the systematics of the measurement are presented in this
  work. In addition, the Telescope Array proton-air inelastic cross
  section  is compared to previous results. Furthermore, the
  proton-proton cross section $\sigma_{\rm p-p}$  is calculated using
  Glauber theory together with the Block, Halzen, and Stanev (BHS)
  QCD-inspired fit
  (~\citep{Block:2011nr},~\citep{Block:2007rq},~\citep{Block05},~\citep{BHS}).
  The proton-proton cross section is also compared to previous $\sigma_{\rm
   p-p}$ experimental results.  Finally, we discuss the summary and the outlook.


\section{Data Trigger, Reconstruction, and Selection.}
\label{sec:data}
\begin{figure}[!h]
  \centering
  \includegraphics[width=0.45\textwidth]{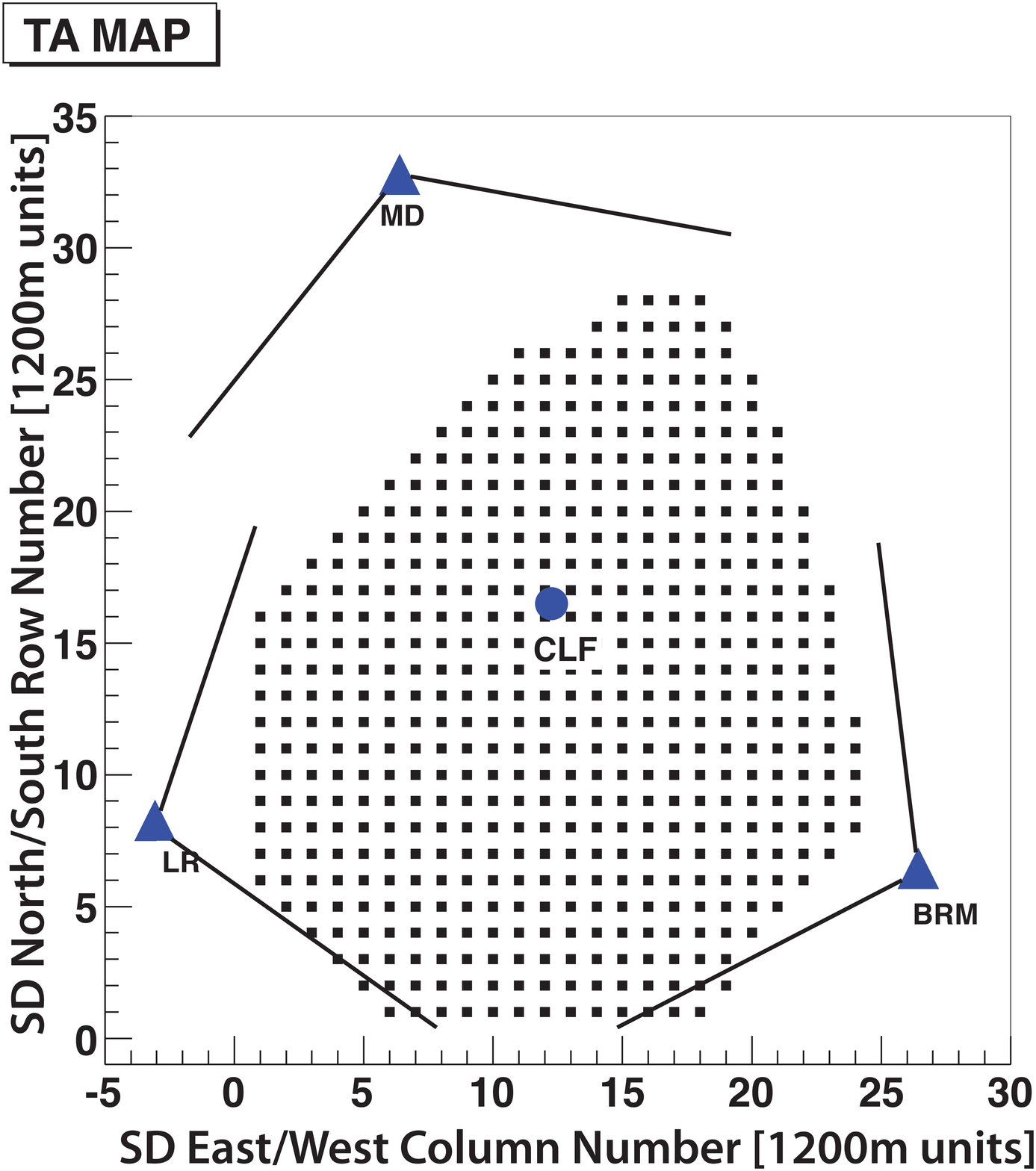}
    \caption{ The Telescope Array detector configuration. The filled squares are the 507 SD scintillators  on a  1.2 km grid. The SD scintillators are enclosed by three fluorescent detectors shown in filled triangles together with their field of view in solid lines. The northernmost  fluorescence detector is called Middle Drum while the southern fluorescence detectors are referred to as Black Rock Mesa and  Long Ridge. The filled circle in the middle equally spaced from the three fluorescence detectors is the Central Laser Facility used for atmospheric monitoring and detector calibration. }
    \label{fig:scheme}
\end{figure}

The data used in this analysis is collected by the Telescope Array (TA) detector located in the 
southwestern desert of the State of Utah. TA is an UHECR detector composed of three Fluorescence Detector (FD) sites and the  Surface Detector array (SD)~\citep{sdnim} as shown in Figure~\ref{fig:scheme}. The SD array occupying $700~{\rm km^{2}}$ and is bounded by the FDs. The northernmost Fluorescence  detector is referred to as  Middle Drum (MD), while the other two southern FDs are named Black Rock Mesa (BRM) and Long Ridge (LR)~\citep{Tokuno}. Moreover, a  Central Laser Facility (CLF) to monitor the atmosphere and calibrate the detector is deployed in the middle of the detector and is located equidistant from the three FDs. 

The two southernmost detectors, LR and BRM, consist of 12 telescopes each, while the  MD detector separated from the northern edge of the SD by 10~km consists of 14 telescopes each of which uses a 5.1~m$^{2}$ spherical mirror. The fluorescence light from each mirror is collected to a camera containing 256 Photo Multiplier Tubes (PMTs) tightly spaced.  Seven of the 14 mirrors view 3$^{\circ}$- 17$^{\circ}$ in elevation while the rest view 17$^{\circ}$- 31$^{\circ}$, with a total azimuth of 112$^{\circ}$ between southwest and southeast. On the other hand, the SD is composed of 507 scintillation counters each 3~m$^{2}$ in area. The SD scintillation counters are spaced on a  1.2~km grid.

The data used in this analysis consists of MD-SD hybrid events. The MD and SD trigger independently. Offline, a hybrid data set is formed by time-matching the events from the two detectors. 
In the monocular mode, an event trigger is recorded
by the SD when three adjacent SDs observe a signal greater than 3
Minimum Ionizing Particles (MIPs) within $8~\mu$s.  When a trigger
occurs the signals from all the SDs within  $\pm~32~\mu$s with
amplitude greater than 0.3 MIP are also recorded. Moreover, a telescope event trigger for the MD is recorded when two subclusters of the 256 PMT cluster triggered within $25~\mu$s. Here a subcluster is defined as a  ($4 \times 4$) 16 PMTs within the 256 PMT cluster. Each subcluster reports a trigger when three tubes in that subcluster trigger within  $25~\mu$s, two of which are adjacent. Finally, multiple  telescope event triggers within $100~\mu$s would be combined into  a single MD event.

Events detected by both detectors (MD and SD)  within $2~\mu$s  are combined into one hybrid event. The combined data set with MD and SD time matched events are then reprocessed using information from both detectors. The FD overlooking the sky above the SD array provides the longitudinal profile of the shower. Meanwhile, the SD provides the event shower core, particle density, and hence improves the geometrical reconstruction significantly. Reference~\cite{Abbasi2014} describes the detector  monocular and hybrid reconstructions of the triggered events in more detail.

To achieve the best $X_{max}$ resolution, a pattern recognition
technique was applied which selected events with a well-defined peak
in the fluorescence light profile. This technique is described more
completely in Reference~\cite{Abbasi2014}. Briefly, each shower
profile's shape was approximated by a set of right triangles, and a
set of cuts on the properties of these triangles was used to reject
$X_{max}$ events with a ``flat'' profile or an indistinct
peak. As shown in Reference~\cite{Abbasi2014}, data to Monte Carlo
comparison studies showed good agreement in basic air shower
distributions such as zenith angle, azimuthal angle, and impact
parameter after these pattern recognition cuts were applied. 

The data used in this analysis is the MD-SD hybrid events collected
between May-2008 and May-2013.  After applying the pattern recognition
cuts to the data we are left with 439 events. The energy range for
this data set is between $10^{18.3}$ and $10^{19.3}$ eV. With an
average energy of $10^{18.68}$ eV, this is equivalent to a 
center of mass energy of $\sqrt s = 95$~TeV. Finally, the $X_{max}$
resolution of this data set achieved after applying the pattern recognition cuts is $\sim 23$~g/cm$^{2}$~\cite{Abbasi2014}.

\section{Analysis}

In this paper we determine the value of $\sigma^{\rm inel}_{\rm p-air}$ using the
$K$-Factor method. This method infers the attenuation length 
and hence the cross section value from the exponential tail
of the  $X_{max}$ distribution. This is assuming that the tail
of the  $X_{max}$ distribution is comprised of the most penetrating/lighter particles (protons).   
The tail of the $X_{max}$ distribution  is fit to the exponential 
$\exp(\frac{-X_{max}}{\Lambda_{m}})$, where $\Lambda_{m}$ is the
attenuation length.  $\Lambda_{m}$ is proportional to the interaction
length  $\lambda_{\rm p-air}$: 
\begin{equation}
\Lambda_{m} = K \lambda_{\rm p-air} =  K \frac{14.45 m_{p}}{\sigma^{\rm
    inel}_{\rm p-air}}
\label{eqn:lamda}
\end{equation}  
where $K$ is dependent on the shower evolution model.
The departure of $K$ from unity depends on the 
pion inelastic cross section and on the inclusive proton and
pion cross sections with the  light nuclear atmospheric target~\citep{Block05}.

In order to determine $K$ to derive the interaction length
$\lambda_{\rm p-air}$ from the slope of $X_{max}$ distribution
$\Lambda_{m}$, we carried out simulation studies using the
one-dimensional air shower Monte Carlo program
CONEX4.37(~\citep{conex},~\citep{conex1},~\citep{conex2}). 
The CONEX program uses a hybrid air shower calculation  for the high energy part of the shower, and
a numerical solution of the cascade equations for the low energy
part of the shower. This hybrid approach of simulating the cosmic
ray showers enables CONEX to be very efficient. Using CONEX allows us
to simulate large number of showers in a very reasonable time scale.  It is worth noting
that the shower parameters obtained with CONEX are consistent with
that obtained with CORISKA~\citep{conex}. 
 
\begin{figure}[!h]
  \centering
  \includegraphics[width=0.50\textwidth]{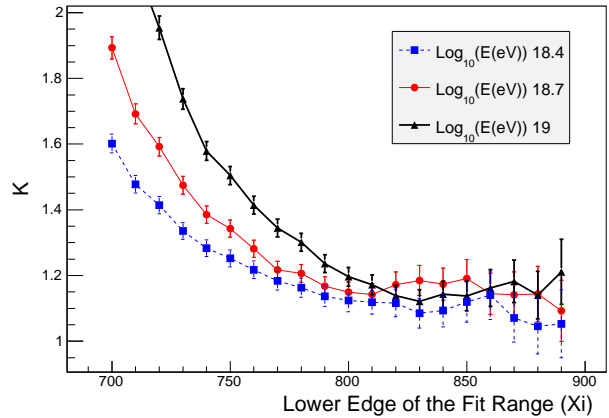}
    \caption{The value of $K$ vs. the lower edge in the fit  range $X_{i}$ to the
      tail of  the $X_{max}$ distribution for several data sets $10^{18.4}, 10^{18.7},$ and
      $10^{19}$ eV simulated using CONEX with the high energy
      model QGSJETII.4. Each data set contains 10,000
      simulated events. }   
    \label{fig:kvalfp}
\end{figure}

Using CONEX  the value of $K$ is determined by simulating 
10,000 events for each of several energy bins for data between $10^{18.3}$ and $10^{19.3}$
eV. The value of $K$ is calculated  for each high energy model for each
energy bin by obtaining the values of ${\Lambda_{m}}$ and
$\lambda_{\rm p-air}$ for that model. The value of ${\Lambda_{m}}$ and
therefore $K$ for
each of the data sets is impacted by the choice of the lower edge of
the fit range $X_{i}$. This dependence is shown in Figure~\ref{fig:kvalfp}.

It is essential that a consistent procedure be used to determine the
$X_{i}$ and consequently the value of $K$  for the shower
simulations and the observed data.  We find from the data that 
$X_{i} = <X_{max}> + 40$~g/cm$^{2}$
is the minimum stable value of $X_{i}$, maximizing the 
number of events in the tail of the distribution and consequently the
statistical power of the measurement.  The same
relative shift distribution is later used in the simulations.
 
\begin{figure}[!h]
  \centering
  \includegraphics[width=0.50\textwidth]{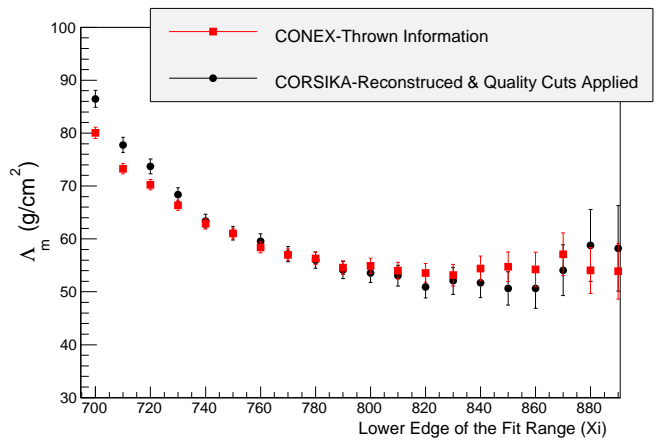}
    \caption{ $\Lambda_{m}$ (g/cm$^2$) vs. the lower edge in the fit  range $X_{i}$ to the
      tail of  the $X_{max}$ distribution at an energy range of
      $10^{18.3}-10^{19.3}$ eV. The value of $\Lambda_{m}$
      is calculated using CONEX with the high energy
      model QGSJETII.4 (square markers). These events were not
      propagated through the detector.
      In addition, the value of $\Lambda_{m}$ is also calculated using
      CORSIKA (circle markers). These events 
      successfully survived the 
      pattern recognition cuts after they were successfully
      detected and  reconstructed.   
    }
    \label{fig:concor}
\end{figure}

It is also important to note that in addition to CONEX we have also
used CORSIKA~\citep{Heck:1998vt}. CORSIKA is used here to simulate
three-dimensional cosmic ray showers. In the simulation process these
showers are thinned in order to reduce the CPU time, and then
dethinned in an attempt to restore lost information~\citep{Stokes}. These showers are then propagated
through the FD and the SD part of the TA detector. The showers that
successfully pass the trigger of the detector are then reconstructed,
after which the pattern recognition event selection are applied. The
value of $\Lambda_{m}$ is then determined and, as shown
in Figure~\ref{fig:concor}, is found
to be consistent with that obtained with CONEX (shower simulation not
propagated through the detector) particularly around the selected
choice of $X_{i} = <X_{max}> + 40$~g/cm$^{2}$. This effect will
also be discussed in Section~\ref{sec:sys}.

\begin{figure}[!h]
  \centering
  \includegraphics[width=0.45\textwidth]{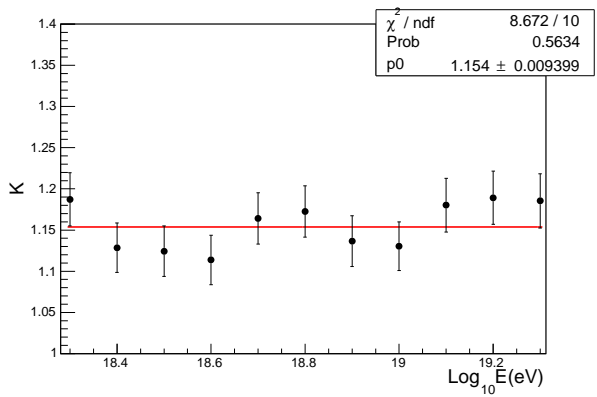}
    \caption{The value of $K$ obtained vs. energy in Log$_{10}$(eV) for  
      simulated data sets using CONEX with the high energy
      model QGSJETII.4, for the energy range of the data,  between  $10^{18.3}$ and 
       $10^{19.3}$ eV.}   
    \label{fig:kvallge}
\end{figure}

The value of $K$  is calculated for each simulated data set between 
the energies of  $10^{18.3}$ and $10^{19.3}$.
Figure~\ref{fig:kvallge} shows $K$ vs. Log$_{10}$(E(eV)).
Note that we have chosen to display QGSJETII.4 as
an example. 
The value of $K$ is then established by fitting the points from Figure~\ref{fig:kvallge}  to a
constant. Table~\ref{tab:kval}  summarizes the high energy models
used, the value of $K$ obtained for these models. It is also worth
mentioning that the K value was also calculated with QGSJETII.3  
and was obtained from this model to be consistent with
that determined from QGSJETII.4 within the statistical fluctuations.
Note that the stability of  $K$ around the
average shown in Figure~\ref{fig:kvallge} 
shows that $K$ is independent of energy and justifies the use of a
single average value over the range of interest.

\begin{figure}[!h]
  \centering
  \includegraphics[width=0.5\textwidth]{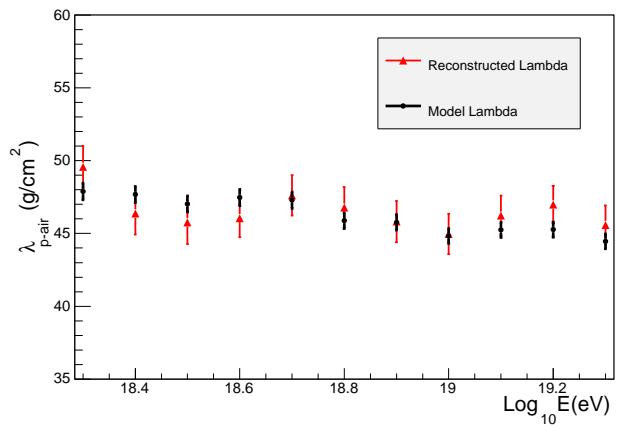}
    \caption{The proton-air interaction length $\lambda_{\rm p-air}$ in g/cm$^{2}$ vs. Energy in
      Log$_{10}$(eV) for the simulated data sets using CONEX with the high energy
      model QGSJETII.4, for the energy range of the data, between $10^{18.3}$ and
      $10^{19.3}$ eV. The circle points are the   $\lambda_{\rm p-air}$
      values obtained from the $X_{1}$ distribution. Triangle points are the ones determined from reconstructing the
      $\lambda_{\rm p-air}$ values using the $K$-Factor method.}   
    \label{fig:lamlge}
\end{figure}

To confirm that the value of $K$ obtained is valid to reproduce the
interaction length of the model, a plot of  $\lambda_{\rm p-air}$
vs. Log$_{10}$(E(eV)) is shown in Figure~\ref{fig:lamlge}. Each point
here represents 10,000 simulated data sets at that energy. The circle
markers are the  $\lambda_{\rm p-air}$ obtained from the $X_{1}$
distributions from the model, while the triangle markers are the
$\lambda_{\rm p-air}$ obtained from reconstruction using the $K$-Factor
method. Figure~\ref{fig:lamlge} shows that using
the  $K$-Factor method does indeed reconstruct the expected values of the
$\lambda_{\rm p-air} $ for these simulations. This ensures that the
value of $K$ obtained in this study describe the value of $K$ of the high energy
models correctly. 

The $K$-Factor determined in the procedure described above is dependent 
on the hadronic interaction model used in the air shower Monte Carlo
simulation.  The high energy models used in this study
are  QGSJETII.4~\citep{qgsjetII}, QGSJET01~\citep{qgsjet01},
SIBYLL~\citep{sibyll}, and EPOS-LHC~\citep{eposlhc}. 
 The resultant values of $K$ determined for these models are summarized
in Table~\ref{tab:kval}.

\begin{center}
  \begin{table}   [!h]
    \begin{tabular}{| p{2cm} | p{2cm} | p{2cm} |}
      \hline
      Model & $K$ \\ \hline
      QGSJETII.4 & 1.15$\pm$ 0.01\\ \hline
      QGSJET01 &  1.22$\pm$0.01 \\ \hline
      SIBYLL &  1.18$\pm$0.01  \\ \hline
      EPOS-LHC &  1.19$\pm$0.01  \\
      \hline
  \end{tabular}
     \caption{ The value of $K$ obtained for each of the high energy
       models. Each $K$ listed is the single average value of $K$
       over the energy range of  $10^{18.3}$-$10^{19.3}$. Note that the values of
       $K$ shows a~$\sim$3$\%$ model uncertainty.}
   \label{tab:kval}%
\end{table}
\end{center}

The first measurement of the proton-air cross section using UHECR was
performed by the Fly’s Eye experiment, which used a calculated value
of $K = 1.6$ and obtained  $\sigma^{inel}_{\rm p-air} = 530 \pm
66$~mb~\cite{FE1987}. Following the Fly's Eye result, the calculated values of
$K$ which appeared in the literature showed a continuous decrease as
full Monte Carlo simulations came into use. By 2000, after the
development of modern high energy hadronic models, the reported
$K$-values still differed by approximately 7\%~\citep{Pryke2000}. Since then, as
shown in Table~\ref{tab:kval}, more complete hadronic shower simulations have
converged on a smaller value of $K = 1.2$, with a model uncertainty of
approximately 3\%. Using this lower K-value, the Fly’s Eye cross
section may be updated to $392 \pm 49$~mb.


\section{Proton-Air Cross Section:}
\label{sec:sys}

\begin{figure}[!h]
  \centering
  \includegraphics[width=0.45\textwidth]{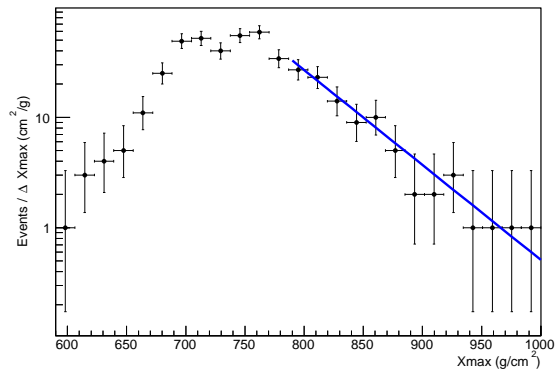}
    \caption{ The number of events per $X_{max}$ bin ($\Delta X_{max}$)  vs. $X_{max}$ g/cm$^{2}$ for the Telescope
      Array data with the energy between $10^{18.3}$ and
      $10^{19.3}$ eV. The line is the exponential fit to the slope.}   
    \label{fig:xmaxdec}
\end{figure}

The data used in this analysis is the Telescope Array Middle Drum-Surface Detector
hybrid events discussed in detail in Section~\ref{sec:data}. 
Figure~\ref{fig:xmaxdec} shows the  $X_{max}$ distribution together with
the exponential unbinned maximum likelihood fit to the tail between
790 and 1000~g/cm$^2$, the
$\Lambda_{m}$ value from the fit is found to be~$(50.47 \pm 6.26 [{\rm Stat.}])$~g/cm$^{2}$.

Consecutively the value of $\sigma^{\rm inel}_{\rm p-air}$ is determined 
where $\sigma^{\rm inel}_{\rm p-air} = K \times 24,160/\Lambda_{m}$ mb using Equation~\ref{eqn:lamda}.  
The $K$ values used are the ones calculated and summarized
 in the previous section in Table~\ref{tab:kval}. Accordingly the
 values of $\sigma^{\rm inel}_{\rm p-air}$ for all the considered hadronic interaction models are determined and  tabulated
 in Table~\ref{tab:sigma}. 
The final value of the proton-air cross inelastic section reported by the
Telescope Array collaboration is the average value of the
$\sigma^{\rm inel}_{\rm p-air}$ 
obtained by the high energy models QGJSETII.4, QGSJET01, SIBYLL, and EPOS-LHC
and is found to be equal to~$(567.0 \pm 70.5 [{\rm Stat.}])$ mb.

\begin{table}  
  \begin{tabular}{| p{3cm} | p{3cm} | }
    \hline
    Model & $\sigma^{\rm inel}_{\rm p-air} \pm [{\rm Stat.}]$  mb   \\ \hline
    QGSJETII.4 & 550.3$\pm$68.5\\ \hline
    QGSJET01 &  583.7$\pm$72.6\\ \hline
    SIBYLL &  564.6$\pm$70.2\\ \hline
    EPOS-LHC &  569.4$\pm$70.8\\
    \hline
  \end{tabular}
  \caption{ The high energy model vs. the $\sigma^{\rm inel}_{\rm p-air}$ in mb
    obtained for that high energy model. }
  \label{tab:sigma}%
\end{table}

In order to quantify the systematic uncertainties on the proton-air cross section obtained using the $K$-Factor method a few different checks were applied.
First,  the systematic value from the hadronic interaction model dependence of the
$\sigma^{\rm inel}_{\rm p-air}$ value  is calculated to be
the maximum difference between the $\sigma^{\rm inel}_{\rm p-air}$  value determined from the
various tested models and the average value obtained from these
models. The systematic uncertainty from the model dependence is
found to be ($\pm 17$) mb.

 In addition, the systematic error in $\sigma^{\rm inel}_{\rm p-air}$
from the systematic error in $\Lambda_{m}$ is also calculated. 
The data is divided in halves based on the zenith angle of the events,
 the distance of the shower using the impact parameter, and finally the energy of the events. 
The  attenuation lengths resulting from all these
subsets are consistent within the statistical fluctuations.
In the case of the energy dependence, below the median of  $10^{18.63}$ eV
$\Lambda_m = 55.7\pm 10.1 $, and above the median $\Lambda_m = 45.5
\pm 7.7$.

Moreover, the systematic effect of
possible energy dependent bias in the  $X_{max}$ distribution was
studied. This is done by shifting the  values of $X_{max}$ by their
elongation rate prior to fitting. The value of $\Lambda_{m}$ is calculated and
the systematic effect from a possible energy bias was found to be negligible.

The next check is calculating the
systematic uncertainty that originates from the detector bias. This includes
the bias that occurs from detecting the events, reconstructing the
events, and applying
the needed cuts to the events. This check is investigated by
comparing the result of the attenuation length $\Lambda_{m}$   of the
simulated shower thrown without any detector effects to the
attenuation length obtained from a three-dimensional shower simulation using
CORSIKA propagated through the detector and reconstructed successfully
including the pattern recognition cuts. As shown in
Figure~\ref{fig:concor}, the value of  $\Lambda_{m}$ was found to be
consistent, for all the high energy
models, between the thrown events and the reconstructed events with
pattern recognition applied. Therefore, the detector bias systematic
effect on the $\Lambda_{m}$ value  is negligible. 

In addition, a fraction of the high energy cosmic rays detected and
used in this study are possibly photons. Such photons may accompany 
the cosmic rays by some scenarios explaining the
origin of ultra high energy cosmic ray sources. In addition, a flux
of photons is also expected from the interaction of cosmic rays with
energies above $4 \times 10^{19} eV $ with the microwave background radiation producing the
Greisen-Zatsepin-Kuz’min (GZK) process (~\citep{Greisen:1966jv},~\citep{Zatsepin:1966jv}).  There have been several
studies placing an upper limit on the integral flux and the fraction
of the primary cosmic ray photons for energies greater than 10$^{18}
$eV
(~\citep{Abu-Zayyad:2013dii},~\citep{Yak2010},~\citep{Abraham:2009qb}). 
 In this study, the lowest derived limit on the photon fraction is
 used and is $< 1\%$~\citep{Settimo}.
 The systematic contribution from the photons is found to be~$+$23 mb.

The result of the proton-air cross section from this work so far
assumes with high energy model simulations a pure protonic cosmic ray composition.
Regardless of what conclusion one would make on the composition 
in the data(~\cite{Abbasi2014},~\cite{augercomp}), the result on the
proton-air cross section from this work would remain the same. However, 
the systematic effect of the presence of other elements in the data beside proton is also
studied. This includes iron, helium, and CNO. Note that the maximum
systematic contribution
from these elements was found to be from helium (deepest $X_{max}$
distribution). Hence, It is the contribution that is
reported in this study. A contribution of~$10\%$,
$20\%$, and $50\%$ from helium and the systematic error associated with such
contribution is reported. For a~$10\%$ contribution the systematic
effect is calculated to be~$- 9$~mb.
Meanwhile, for a~$20\%$ and $50\%$ contribution the systematic effect is
determined to be~$- 18$ mb and $-42$ mb respectively. The final
systematic value, conservatively assuming a
$20 \%$ Helium contamination, is calculated by adding in quadrature
the systematic values and is found to be ($-$25,$+$29) mb.  Table~\ref{tab:sys} summarizes the systematic checks for the
 proton-air cross section, including the final systematic value.

 \begin{table}  
   \begin{tabular}{| p{3cm} | p{3cm} | }
     \hline
     Systematic source & Systematics (mb)   \\ \hline
     Model Dependence & ($\pm$17)  \\ \hline
     $10\%$ Helium &  $-$ 9\\ \hline
     $20\%$ Helium &  $-$ 18\\ \hline
     $50\%$ Helium &  $-$ 42\\ \hline
     Gamma &  $+$ 23  \\ \hline
     Summary    &  ($-$25,$+$29) \\
     ($20\%$ Helium)  &  \\
     \hline
   \end{tabular}
   \caption{The systematic source vs. The systematic values of that source.}
 \label{tab:sys}%
 \end{table}

We summarize the result of the our proton-air cross section obtained
using the $K$-Factor method described previously together with the
systematic checks obtained to be equal to 

\begin{equation}
\sigma^{\rm inel}_{\rm p-air} = 567.0 \pm 70.5 [{\rm Stat.}]  ^{+29}_{-25} [{\rm Sys.}]
{\rm mb}. 
\end{equation}

\begin{figure}[!h]
  \centering
 \includegraphics[width=0.5\textwidth]{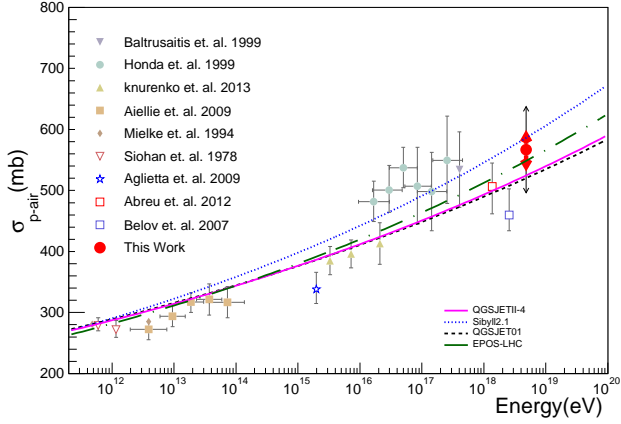}
    \caption{ The proton-air cross section result of this work,
      including the statistical (outer/thinner) and systematic (inner/thicker) error bar. The result of this work is shown  in
      comparison to other experimental results (~\citep{Siohan:1978zk},~\citep{FE1987},~\citep{Honda1992},~\citep{Knurenko2013},~\citep{Mielke:1994un},~\citep{Belov2006},~\citep{Aielli:2009ca},~\citep{Auger2012},~\citep{EAStop}). In addition, 
      the high energy models (QGSJETII.4, QGSJET01, SIBYLL, EPOS-LHC)  cross
      section predictions are also shown by solid line, fine dashed line,
      dotted line, and dashed line consecutively.}   
    \label{fig:pair}
\end{figure}

This is obtained at an average energy of~10$^{18.68}$eV.  The result of the proton-air cross section
is then compared to the results obtained from various experimental
results
(~\citep{Siohan:1978zk},~\citep{FE1987},~\citep{Honda1992},~\citep{Knurenko2013},~\citep{Mielke:1994un},~\citep{Belov2006},~\citep{Aielli:2009ca},~\citep{Auger2012},~\citep{EAStop})~Figures~\ref{fig:pair}. In addition,  the
experimental  results of the high energy models (QGSJETII.4, QGSJET01,
SIBYLL, EPOS-LHC) cross section predictions are also included. This includes the statistical (outer/thinner error bar) and the systematic (inner/thicker error bar).

\section{Proton-Proton Cross section}

From the TA proton-air cross section result we can determine the total
proton-proton cross section. The process of inferring $\sigma_{p-p}$
from $\sigma^{\rm inel}_{p-air}$ is described in details in
~\citep{Engel:1998pw}, and ~\citep{Gaisser:1986haa}. 

The $\sigma_{p-p}$ is calculated from the measured cross section,
also known as the inelastic cross section $\sigma_{p-air}^{\rm inel}$, using
both Glauber Formalism~\citep{Glauber70} and the relation:
\begin{equation}
\sigma_{p-air}^{\rm inel} = \sigma_{p-air}^{\rm total} -
\sigma_{p-air}^{\rm el} - \sigma_{p-air}^{\rm qel}
\label{eqn:inelatot}
\end{equation}
Where $\sigma_{p-air}^{\rm total}$  is the total cross section,
$\sigma_{p-air}^{\rm el}$ is the elastic cross section and
$\sigma_{p-air}^{\rm qel}$ is the quasi elastic cross section. 
The quasi-elastic cross section corresponds to scattering processes in
which nuclear excitation occurs without particle production.

The relation between the $\sigma_{p-air}^{\rm inel}$ and the
$\sigma_{\rm p-p}$ is
highly dependent on the forward scattering elastic slope $B$.

\begin{equation}
B =  \frac{d}{dt} \left[ \ln \frac{d\sigma_{\rm p-p}^{\rm el}}{dt} \right]_{t=0} 
\label{eqn:bsigma}
\end{equation}  

\begin{figure}[!h]
  \centering
  \includegraphics[width=0.45\textwidth]{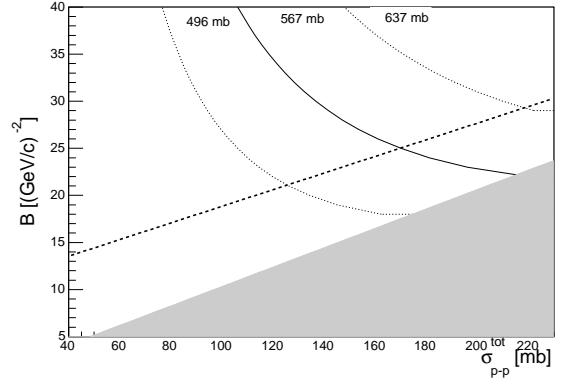}
    \caption{ The elastic slope $B$ in
      ((GeV/c)$^{-2}$) vs. $\sigma_{\rm p-p}^{\rm total}$ in mb. The solid and the
      dotted curves are the relation between $B$ and
      $\sigma_{\rm p-p}^{\rm total}$ for the constant value of the measured
      $\sigma_{\rm p-air}^{\rm inel}$ by the Telescope Array detector and the
      statistical error using Glauber Formalism. The dashed line is the BHS  QCD inspired fit~\citep{BHS}. While the gray shaded area is the unitarity constraint.  }   
    \label{fig:bsigma}
\end{figure}

This is
shown in the $B$, $\sigma_{p-p}^{\rm total}$ plane in
Figure~\ref{fig:bsigma}. Here the solid and dotted curves represent a
constant value of  $\sigma^{\rm inel}_{\rm p-air}$ that reflects the Telescope Array
measured value and the statistical fluctuations. 

 There have been many theories predicting the relationship between $B$ and
 $\sigma_{\rm p-p}$. However many of these models either failed to describe
 the elastic scattering data, or the elastic slope energy dependence
 from the Tevatron
 (~\citep{Engel:1998pw},~\citep{DiasdeDeus1978},~\citep{Buras:1973km}). A more updated
 theory using the single pomeron exchange model while describing the
Tevatron  data correctly is not consistent with the Unitarity
constraint (~\citep{Engel:1998pw},~\citep{Donnachie:1992ny}). Here the
unitarity constraint is shown by solid grey shaded area in  Figure~\ref{fig:bsigma}. A more recent prediction is the Block, Halzen, and Stanev
(BHS) fit~\citep{Block:2011nr}. It is consistent with unitarity while 
using a QCD inspired fit to the pp and $\bar{\rm p}$p data from the
Tevatron. The dashed line in Figure~\ref{fig:bsigma} shows the BHS
prediction. Here $\sigma_{p-air}^{\rm inel}$ is converted to
$\sigma_{p-p}^{\rm total}$
using the BHS fit. The statistical and systematic errors in 
$\sigma_{\rm p-p}^{\rm total}$ are propagated from the $\sigma_{\rm p-air}^{\rm inel}$ statistical and systematic error calculation.
 The  $\sigma_{\rm p-p}^{\rm total}$ is found to be $170_{-44}^{+48} [{\rm Stat.}] _{-17}^{+19} [{\rm Sys.}]$ mb. 

\begin{figure}[!h]
  \centering
  \includegraphics[width=0.45\textwidth]{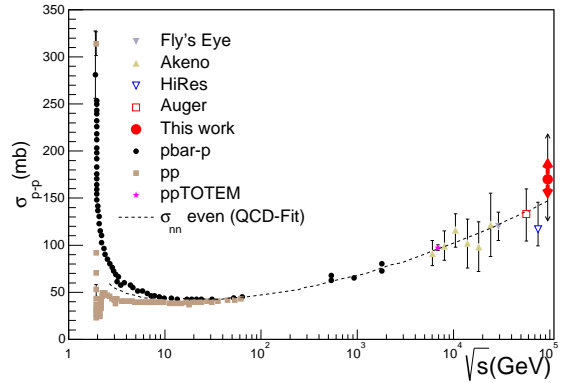}
    \caption{ The proton-proton cross section vs. the 
      center of mass energy result of this work,
      including the statistical (outer/thinner) and systematic (inner/thicker) error
      bars.  The $\bar{\rm p}$p and the pp data are shown in 
      smaller darker circles and square symbols 
      consecutively~\citep{pp1999}. The recent result from
     LHC is also shown by the star marker~\citep{Totem2011}. The result of this work is shown in comparison
      to previous work by cosmic rays
      detectors(~\citep{FE1987},~\citep{Honda1992},~\citep{Belov2006},~\citep{Auger2012}). The dashed curve is the QCD inspired fit by BHS~\citep{Block05}.}
    \label{fig:ppcrosssec}
\end{figure}

The $\sigma_{\rm p-p}^{total}$ calculated in this work is shown in 
Figures~\ref{fig:ppcrosssec} compared to
previous results from cosmic ray experiments like Fly's Eye~\citep{FE1987},
Akeno~\citep{Honda1992}, HiRes~\citep{Belov2006}, and Auger~\citep{Auger2012}, together with
accelerator pp and $\bar{\rm p}$p cross section
measurement~\citep{pp1999}, in addition to the recent result from LHC
by TOTEM~\citep{Totem2011}. The dotted
curve is  the QCD inspired fit of the total p-p cross section vs. the
center of mass energy $\sqrt{s}$(GeV)~\citep{Block05}. The result from
this work at $\sqrt{s}$ = $95_{-8}^{+5}$ TeV, in addition to the most recent result published by the
Auger experiment at $\sqrt{s}$ = 57 TeV~\citep{Auger2012}
and reported recent result by the LHC at $\sqrt{s}$ = 7
TeV~\citep{Totem2011} are all in agreement
with the fit.   

\section{Conclusion and Outlook}

In this work we used events collected by Telescope Array between May-2008 and May-2013 in hybrid mode to determine the $\sigma^{\rm inel}_{\rm p-air}$ using the $K$-Factor method. The hadronic model dependence of the  $K$-Factor method  was investigated. The latest updated hadronic interaction models have converged with time on the value of $K$ with an uncertainty of~$\sim$3$\%$. This makes the $K$-Factor method a weakly model dependent method to use in calculating the $\sigma^{\rm inel}_{\rm p-air}$. Several systematic checks were applied and the final value of  $\sigma^{\rm inel}_{\rm p-air}$ was found to be equal to~$567.0 \pm 70.5 [{\rm Stat.}]  ^{+29}_{-25} [{\rm Sys.}]$ mb.

Ultimately the value of $\sigma_{\rm p-p}$  is determined  from  $\sigma^{\rm inel}_{\rm p-air}$ using Glauber theory and BHS QCD inspired fit. Such a fundamental  measurement  at this high energy ($\sqrt{s} = 95_{-8}^{+5}$ TeV) could not be obtained with  current particle accelerators.The value of $\sigma^{\rm tot}_{\rm p-p}$ was determined to be $170
_{-44}^{+48} [{\rm Stat.}]  _{-17}^{+19} [{\rm Sys.}] $ mb. 

While the events used in this analysis were collected with MD-SD part of the detector, future cross section results, using 
thoroughly  analysed events could be performed using LR and BRM data, and ultimately with the full detector.  LR and BRM are the fluorescence detectors closer in distance to the SD and therefore we could extend the energy range of the collected data down to~1~EeV . This will enable us to study the measurement down to ~1~EeV with higher statistical power which would allow us to constrain the available high energy model cross section predictions.  
\section{Acknowledgements}
Many thanks to CONEX and CORSIKA authors. Particular thanks for 
Tanguy Pierog for his help. The Telescope Array experiment is supported by the Japan
Society for the Promotion of Science through Grants-in-Aid for Scientific Research on Specially 
Promoted Research (21000002) ``Extreme Phenomena in the Universe Explored
by Highest Energy Cosmic Rays'' and for Scientific Research
 (19104006), and the Inter-University Research Program of
the Institute for Cosmic Ray Research; by the U.S. National
Science Foundation awards PHY-0307098, PHY-0601915, PHY-0649681,
PHY-0703893, PHY-0758342, PHY-0848320, PHY-1069280, PHY-1069286, PHY-1404495 and PHY-1404502; by the National Research Foundation
of Korea (2007-0093860, R32-10130, 2012R1A1A2008381, 2013004883); by
the Russian Academy of Sciences, RFBR grants 11-02-01528a and 13-02-01311a (INR), IISN project No. 4.4502.13, and Belgian
Science Policy under IUAP VII/37 (ULB). The foundations
of Dr. Ezekiel R. and Edna Wattis Dumke, Willard
L. Eccles, and George S. and Dolores Dor\'e Eccles all
helped with generous donations. The State of Utah supported
the project through its Economic Development Board, and the
University of Utah through the Office of the Vice President
for Research. The experimental site became available through
the cooperation of the Utah School and Institutional Trust
Lands Administration (SITLA), U.S. Bureau of Land Management,
and the U.S. Air Force. We also wish to thank the people
and the officials of Millard County, Utah for their steadfast
and warm support. We gratefully acknowledge the contributions
from the technical staffs of our home institutions. An
allocation of computer time from the Center for High Performance
Computing at the University of Utah is gratefully acknowledged.


\begin{thebibliography}{42}%
\makeatletter
\providecommand \@ifxundefined [1]{%
 \@ifx{#1\undefined}
}%
\providecommand \@ifnum [1]{%
 \ifnum #1\expandafter \@firstoftwo
 \else \expandafter \@secondoftwo
 \fi
}%
\providecommand \@ifx [1]{%
 \ifx #1\expandafter \@firstoftwo
 \else \expandafter \@secondoftwo
 \fi
}%
\providecommand \natexlab [1]{#1}%
\providecommand \enquote  [1]{``#1''}%
\providecommand \bibnamefont  [1]{#1}%
\providecommand \bibfnamefont [1]{#1}%
\providecommand \citenamefont [1]{#1}%
\providecommand \href@noop [0]{\@secondoftwo}%
\providecommand \href [0]{\begingroup \@sanitize@url \@href}%
\providecommand \@href[1]{\@@startlink{#1}\@@href}%
\providecommand \@@href[1]{\endgroup#1\@@endlink}%
\providecommand \@sanitize@url [0]{\catcode `\\12\catcode `\$12\catcode
  `\&12\catcode `\#12\catcode `\^12\catcode `\_12\catcode `\%12\relax}%
\providecommand \@@startlink[1]{}%
\providecommand \@@endlink[0]{}%
\providecommand \url  [0]{\begingroup\@sanitize@url \@url }%
\providecommand \@url [1]{\endgroup\@href {#1}{\urlprefix }}%
\providecommand \urlprefix  [0]{URL }%
\providecommand \Eprint [0]{\href }%
\providecommand \doibase [0]{http://dx.doi.org/}%
\providecommand \selectlanguage [0]{\@gobble}%
\providecommand \bibinfo  [0]{\@secondoftwo}%
\providecommand \bibfield  [0]{\@secondoftwo}%
\providecommand \translation [1]{[#1]}%
\providecommand \BibitemOpen [0]{}%
\providecommand \bibitemStop [0]{}%
\providecommand \bibitemNoStop [0]{.\EOS\space}%
\providecommand \EOS [0]{\spacefactor3000\relax}%
\providecommand \BibitemShut  [1]{\csname bibitem#1\endcsname}%
\let\auto@bib@innerbib\@empty
\bibitem [{\citenamefont {Baltrusaitis}\ \emph {et~al.}(1984)\citenamefont
  {Baltrusaitis}, \citenamefont {Cassiday}, \citenamefont {Elbert},
  \citenamefont {Gerhardy}, \citenamefont {Ko} \emph {et~al.}}]{FE1987}%
  \BibitemOpen
  \bibfield  {author} {\bibinfo {author} {\bibfnamefont {R.}~\bibnamefont
  {Baltrusaitis}}, \bibinfo {author} {\bibfnamefont {G.}~\bibnamefont
  {Cassiday}}, \bibinfo {author} {\bibfnamefont {J.}~\bibnamefont {Elbert}},
  \bibinfo {author} {\bibfnamefont {P.}~\bibnamefont {Gerhardy}}, \bibinfo
  {author} {\bibfnamefont {S.}~\bibnamefont {Ko}},  \emph {et~al.},\ }\href
  {\doibase 10.1103/PhysRevLett.52.1380} {\bibfield  {journal} {\bibinfo
  {journal} {Phys.Rev.Lett.}\ }\textbf {\bibinfo {volume} {52}},\ \bibinfo
  {pages} {1380} (\bibinfo {year} {1984})}\BibitemShut {NoStop}%
\bibitem [{\citenamefont {Abreu}\ \emph {et~al.}(2012)\citenamefont {Abreu}
  \emph {et~al.}}]{Auger2012}%
  \BibitemOpen
  \bibfield  {author} {\bibinfo {author} {\bibfnamefont {P.}~\bibnamefont
  {Abreu}} \emph {et~al.} (\bibinfo {collaboration} {Pierre Auger
  Collaboration}),\ }\href {\doibase 10.1103/PhysRevLett.109.062002} {\bibfield
   {journal} {\bibinfo  {journal} {Phys.Rev.Lett.}\ }\textbf {\bibinfo {volume}
  {109}},\ \bibinfo {pages} {062002} (\bibinfo {year} {2012})},\ \Eprint
  {http://arxiv.org/abs/1208.1520} {arXiv:1208.1520 [hep-ex]} \BibitemShut
  {NoStop}%
\bibitem [{\citenamefont {Ulrich}\ \emph {et~al.}(2011)\citenamefont {Ulrich},
  \citenamefont {Engel},\ and\ \citenamefont {Unger}}]{Ulrich2010}%
  \BibitemOpen
  \bibfield  {author} {\bibinfo {author} {\bibfnamefont {R.}~\bibnamefont
  {Ulrich}}, \bibinfo {author} {\bibfnamefont {R.}~\bibnamefont {Engel}}, \
  and\ \bibinfo {author} {\bibfnamefont {M.}~\bibnamefont {Unger}},\ }\href
  {\doibase 10.1103/PhysRevD.83.054026} {\bibfield  {journal} {\bibinfo
  {journal} {Phys.Rev.}\ }\textbf {\bibinfo {volume} {D83}},\ \bibinfo {pages}
  {054026} (\bibinfo {year} {2011})},\ \Eprint {http://arxiv.org/abs/1010.4310}
  {arXiv:1010.4310 [hep-ph]} \BibitemShut {NoStop}%
\bibitem [{\citenamefont {Abbasi}\ \emph {et~al.}(2014)\citenamefont {Abbasi},
  \citenamefont {Abe}, \citenamefont {Abu-Zayyad}, \citenamefont {Allen},
  \citenamefont {Anderson} \emph {et~al.}}]{Abbasi2014}%
  \BibitemOpen
  \bibfield  {author} {\bibinfo {author} {\bibfnamefont {R.}~\bibnamefont
  {Abbasi}}, \bibinfo {author} {\bibfnamefont {M.}~\bibnamefont {Abe}},
  \bibinfo {author} {\bibfnamefont {T.}~\bibnamefont {Abu-Zayyad}}, \bibinfo
  {author} {\bibfnamefont {M.}~\bibnamefont {Allen}}, \bibinfo {author}
  {\bibfnamefont {R.}~\bibnamefont {Anderson}},  \emph {et~al.},\ }\href
  {\doibase 10.1016/j.astropartphys.2014.11.004} {\bibfield  {journal}
  {\bibinfo  {journal} {Astropart.Phys.}\ }\textbf {\bibinfo {volume} {64}},\
  \bibinfo {pages} {49} (\bibinfo {year} {2014})},\ \Eprint
  {http://arxiv.org/abs/1408.1726} {arXiv:1408.1726 [astro-ph.HE]} \BibitemShut
  {NoStop}%
\bibitem [{\citenamefont {Block}(2011)}]{Block:2011nr}%
  \BibitemOpen
  \bibfield  {author} {\bibinfo {author} {\bibfnamefont {M.}~\bibnamefont
  {Block}},\ }\href {\doibase 10.1103/PhysRevD.84.091501} {\bibfield  {journal}
  {\bibinfo  {journal} {Phys.Rev.}\ }\textbf {\bibinfo {volume} {D84}},\
  \bibinfo {pages} {091501} (\bibinfo {year} {2011})},\ \Eprint
  {http://arxiv.org/abs/1109.2940} {arXiv:1109.2940 [hep-ph]} \BibitemShut
  {NoStop}%
\bibitem [{\citenamefont {Block}(2007)}]{Block:2007rq}%
  \BibitemOpen
  \bibfield  {author} {\bibinfo {author} {\bibfnamefont {M.~M.}\ \bibnamefont
  {Block}},\ }\href {\doibase 10.1103/PhysRevD.76.111503} {\bibfield  {journal}
  {\bibinfo  {journal} {Phys. Rev.}\ }\textbf {\bibinfo {volume} {D76}},\
  \bibinfo {pages} {111503} (\bibinfo {year} {2007})},\ \Eprint
  {http://arxiv.org/abs/0705.3037} {arXiv:0705.3037 [hep-ph]} \BibitemShut
  {NoStop}%
\bibitem [{\citenamefont {Block}\ and\ \citenamefont {Halzen}(2005)}]{Block05}%
  \BibitemOpen
  \bibfield  {author} {\bibinfo {author} {\bibfnamefont {M.}~\bibnamefont
  {Block}}\ and\ \bibinfo {author} {\bibfnamefont {F.}~\bibnamefont {Halzen}},\
  }\href {\doibase 10.1103/PhysRevD.72.036006, 10.1103/PhysRevD.72.039902,
  10.1103/PhysRevD.72.036006 10.1103/PhysRevD.72.039902} {\bibfield  {journal}
  {\bibinfo  {journal} {Phys.Rev.}\ }\textbf {\bibinfo {volume} {D72}},\
  \bibinfo {pages} {036006} (\bibinfo {year} {2005})},\ \Eprint
  {http://arxiv.org/abs/hep-ph/0506031} {arXiv:hep-ph/0506031 [hep-ph]}
  \BibitemShut {NoStop}%
\bibitem [{\citenamefont {Block}\ \emph {et~al.}(2000)\citenamefont {Block},
  \citenamefont {Halzen},\ and\ \citenamefont {Stanev}}]{BHS}%
  \BibitemOpen
  \bibfield  {author} {\bibinfo {author} {\bibfnamefont {M.}~\bibnamefont
  {Block}}, \bibinfo {author} {\bibfnamefont {F.}~\bibnamefont {Halzen}}, \
  and\ \bibinfo {author} {\bibfnamefont {T.}~\bibnamefont {Stanev}},\ }\href
  {\doibase 10.1103/PhysRevD.62.077501} {\bibfield  {journal} {\bibinfo
  {journal} {Phys.Rev.}\ }\textbf {\bibinfo {volume} {D62}},\ \bibinfo {pages}
  {077501} (\bibinfo {year} {2000})},\ \Eprint
  {http://arxiv.org/abs/hep-ph/0004232} {arXiv:hep-ph/0004232 [hep-ph]}
  \BibitemShut {NoStop}%
\bibitem [{\citenamefont {Abu-Zayyad}\ \emph {et~al.}(2012)\citenamefont
  {Abu-Zayyad} \emph {et~al.}}]{sdnim}%
  \BibitemOpen
  \bibfield  {author} {\bibinfo {author} {\bibfnamefont {T.}~\bibnamefont
  {Abu-Zayyad}} \emph {et~al.} (\bibinfo {collaboration} {Telescope Array
  Collaboration}),\ }\href {\doibase 10.1016/j.nima.2012.05.079} {\bibfield
  {journal} {\bibinfo  {journal} {Nucl.Instrum.Meth.}\ }\textbf {\bibinfo
  {volume} {A689}},\ \bibinfo {pages} {87} (\bibinfo {year} {2012})},\ \Eprint
  {http://arxiv.org/abs/1201.4964} {arXiv:1201.4964 [astro-ph.IM]} \BibitemShut
  {NoStop}%
\bibitem [{\citenamefont {Tokuno}\ \emph {et~al.}(2012)\citenamefont {Tokuno},
  \citenamefont {Tameda}, \citenamefont {Takeda}, \citenamefont {Kadota},
  \citenamefont {Ikeda} \emph {et~al.}}]{Tokuno}%
  \BibitemOpen
  \bibfield  {author} {\bibinfo {author} {\bibfnamefont {H.}~\bibnamefont
  {Tokuno}}, \bibinfo {author} {\bibfnamefont {Y.}~\bibnamefont {Tameda}},
  \bibinfo {author} {\bibfnamefont {M.}~\bibnamefont {Takeda}}, \bibinfo
  {author} {\bibfnamefont {K.}~\bibnamefont {Kadota}}, \bibinfo {author}
  {\bibfnamefont {D.}~\bibnamefont {Ikeda}},  \emph {et~al.},\ }\href {\doibase
  10.1016/j.nima.2012.02.044} {\bibfield  {journal} {\bibinfo  {journal}
  {Nucl.Instrum.Meth.}\ }\textbf {\bibinfo {volume} {A676}},\ \bibinfo {pages}
  {54} (\bibinfo {year} {2012})},\ \Eprint {http://arxiv.org/abs/1201.0002}
  {arXiv:1201.0002 [astro-ph.IM]} \BibitemShut {NoStop}%
\bibitem [{\citenamefont {Bergmann}\ \emph {et~al.}(2007)\citenamefont
  {Bergmann}, \citenamefont {Engel}, \citenamefont {Heck}, \citenamefont
  {Kalmykov}, \citenamefont {Ostapchenko} \emph {et~al.}}]{conex}%
  \BibitemOpen
  \bibfield  {author} {\bibinfo {author} {\bibfnamefont {T.}~\bibnamefont
  {Bergmann}}, \bibinfo {author} {\bibfnamefont {R.}~\bibnamefont {Engel}},
  \bibinfo {author} {\bibfnamefont {D.}~\bibnamefont {Heck}}, \bibinfo {author}
  {\bibfnamefont {N.}~\bibnamefont {Kalmykov}}, \bibinfo {author}
  {\bibfnamefont {S.}~\bibnamefont {Ostapchenko}},  \emph {et~al.},\ }\href
  {\doibase 10.1016/j.astropartphys.2006.08.005} {\bibfield  {journal}
  {\bibinfo  {journal} {Astropart.Phys.}\ }\textbf {\bibinfo {volume} {26}},\
  \bibinfo {pages} {420} (\bibinfo {year} {2007})},\ \Eprint
  {http://arxiv.org/abs/astro-ph/0606564} {arXiv:astro-ph/0606564 [astro-ph]}
  \BibitemShut {NoStop}%
\bibitem [{\citenamefont {Pierog}\ \emph {et~al.}(2006)\citenamefont {Pierog}
  \emph {et~al.}}]{conex1}%
  \BibitemOpen
  \bibfield  {author} {\bibinfo {author} {\bibfnamefont {T.}~\bibnamefont
  {Pierog}} \emph {et~al.},\ }\href@noop {} {\bibfield  {journal} {\bibinfo
  {journal} {Nucl. Phys. Proc. Suppl.}\ }\textbf {\bibinfo {volume} {151}},\
  \bibinfo {pages} {159} (\bibinfo {year} {2006})},\ \Eprint
  {http://arxiv.org/abs/astro-ph/0411260} {astro-ph/0411260} \BibitemShut
  {NoStop}%
\bibitem [{\citenamefont {Bossard}\ \emph {et~al.}(2001)\citenamefont {Bossard}
  \emph {et~al.}}]{conex2}%
  \BibitemOpen
  \bibfield  {author} {\bibinfo {author} {\bibfnamefont {G.}~\bibnamefont
  {Bossard}} \emph {et~al.},\ }\href@noop {} {\bibfield  {journal} {\bibinfo
  {journal} {Phys. Rev.}\ }\textbf {\bibinfo {volume} {D63}},\ \bibinfo {pages}
  {054030} (\bibinfo {year} {2001})},\ \Eprint
  {http://arxiv.org/abs/hep-ph/0009119} {hep-ph/0009119} \BibitemShut {NoStop}%
\bibitem [{\citenamefont {Heck}\ \emph {et~al.}(1998)\citenamefont {Heck},
  \citenamefont {Schatz}, \citenamefont {Thouw}, \citenamefont {Knapp},\ and\
  \citenamefont {Capdevielle}}]{Heck:1998vt}%
  \BibitemOpen
  \bibfield  {author} {\bibinfo {author} {\bibfnamefont {D.}~\bibnamefont
  {Heck}}, \bibinfo {author} {\bibfnamefont {G.}~\bibnamefont {Schatz}},
  \bibinfo {author} {\bibfnamefont {T.}~\bibnamefont {Thouw}}, \bibinfo
  {author} {\bibfnamefont {J.}~\bibnamefont {Knapp}}, \ and\ \bibinfo {author}
  {\bibfnamefont {J.}~\bibnamefont {Capdevielle}},\ }\href@noop {} {\
  (\bibinfo {year} {1998})}\BibitemShut {NoStop}%
\bibitem [{\citenamefont {Stokes}\ \emph {et~al.}(2012)\citenamefont {Stokes},
  \citenamefont {Cady}, \citenamefont {Ivanov}, \citenamefont {Matthews},\ and\
  \citenamefont {Thomson}}]{Stokes}%
  \BibitemOpen
  \bibfield  {author} {\bibinfo {author} {\bibfnamefont {B.}~\bibnamefont
  {Stokes}}, \bibinfo {author} {\bibfnamefont {R.}~\bibnamefont {Cady}},
  \bibinfo {author} {\bibfnamefont {D.}~\bibnamefont {Ivanov}}, \bibinfo
  {author} {\bibfnamefont {J.}~\bibnamefont {Matthews}}, \ and\ \bibinfo
  {author} {\bibfnamefont {G.}~\bibnamefont {Thomson}},\ }\href {\doibase
  10.1016/j.astropartphys.2012.03.004} {\bibfield  {journal} {\bibinfo
  {journal} {Astropart.Phys.}\ }\textbf {\bibinfo {volume} {35}},\ \bibinfo
  {pages} {759} (\bibinfo {year} {2012})},\ \Eprint
  {http://arxiv.org/abs/1104.3182} {arXiv:1104.3182 [astro-ph.IM]} \BibitemShut
  {NoStop}%
\bibitem [{\citenamefont {Ostapchenko}(2006)}]{qgsjetII}%
  \BibitemOpen
  \bibfield  {author} {\bibinfo {author} {\bibfnamefont {S.}~\bibnamefont
  {Ostapchenko}},\ }\href {\doibase 10.1016/j.nuclphysbps.2005.07.026}
  {\bibfield  {journal} {\bibinfo  {journal} {Nucl.Phys.Proc.Suppl.}\ }\textbf
  {\bibinfo {volume} {151}},\ \bibinfo {pages} {143} (\bibinfo {year}
  {2006})},\ \Eprint {http://arxiv.org/abs/hep-ph/0412332}
  {arXiv:hep-ph/0412332 [hep-ph]} \BibitemShut {NoStop}%
\bibitem [{\citenamefont {Kalmykov}\ \emph {et~al.}(1997)\citenamefont
  {Kalmykov}, \citenamefont {Ostapchenko},\ and\ \citenamefont
  {Pavlov}}]{qgsjet01}%
  \BibitemOpen
  \bibfield  {author} {\bibinfo {author} {\bibfnamefont {N.}~\bibnamefont
  {Kalmykov}}, \bibinfo {author} {\bibfnamefont {S.}~\bibnamefont
  {Ostapchenko}}, \ and\ \bibinfo {author} {\bibfnamefont {A.}~\bibnamefont
  {Pavlov}},\ }\href {\doibase 10.1016/S0920-5632(96)00846-8} {\bibfield
  {journal} {\bibinfo  {journal} {Nucl.Phys.Proc.Suppl.}\ }\textbf {\bibinfo
  {volume} {52B}},\ \bibinfo {pages} {17} (\bibinfo {year} {1997})}\BibitemShut
  {NoStop}%
\bibitem [{\citenamefont {Ahn}\ \emph {et~al.}(2009)\citenamefont {Ahn},
  \citenamefont {Engel}, \citenamefont {Gaisser}, \citenamefont {Lipari},\ and\
  \citenamefont {Stanev}}]{sibyll}%
  \BibitemOpen
  \bibfield  {author} {\bibinfo {author} {\bibfnamefont {E.-J.}\ \bibnamefont
  {Ahn}}, \bibinfo {author} {\bibfnamefont {R.}~\bibnamefont {Engel}}, \bibinfo
  {author} {\bibfnamefont {T.~K.}\ \bibnamefont {Gaisser}}, \bibinfo {author}
  {\bibfnamefont {P.}~\bibnamefont {Lipari}}, \ and\ \bibinfo {author}
  {\bibfnamefont {T.}~\bibnamefont {Stanev}},\ }\href {\doibase
  10.1103/PhysRevD.80.094003} {\bibfield  {journal} {\bibinfo  {journal}
  {Phys.Rev.}\ }\textbf {\bibinfo {volume} {D80}},\ \bibinfo {pages} {094003}
  (\bibinfo {year} {2009})},\ \Eprint {http://arxiv.org/abs/0906.4113}
  {arXiv:0906.4113 [hep-ph]} \BibitemShut {NoStop}%
\bibitem [{\citenamefont {Pierog}\ \emph {et~al.}(2013)\citenamefont {Pierog},
  \citenamefont {Karpenko}, \citenamefont {Katzy}, \citenamefont {Yatsenko},\
  and\ \citenamefont {Werner}}]{eposlhc}%
  \BibitemOpen
  \bibfield  {author} {\bibinfo {author} {\bibfnamefont {T.}~\bibnamefont
  {Pierog}}, \bibinfo {author} {\bibfnamefont {I.}~\bibnamefont {Karpenko}},
  \bibinfo {author} {\bibfnamefont {J.}~\bibnamefont {Katzy}}, \bibinfo
  {author} {\bibfnamefont {E.}~\bibnamefont {Yatsenko}}, \ and\ \bibinfo
  {author} {\bibfnamefont {K.}~\bibnamefont {Werner}},\ }\href@noop {} {\
  (\bibinfo {year} {2013})},\ \Eprint {http://arxiv.org/abs/1306.0121}
  {arXiv:1306.0121 [hep-ph]} \BibitemShut {NoStop}%
\bibitem [{\citenamefont {Pryke}(2001)}]{Pryke2000}%
  \BibitemOpen
  \bibfield  {author} {\bibinfo {author} {\bibfnamefont {C.}~\bibnamefont
  {Pryke}},\ }\href {\doibase 10.1016/S0927-6505(00)00132-8} {\bibfield
  {journal} {\bibinfo  {journal} {Astropart.Phys.}\ }\textbf {\bibinfo {volume}
  {14}},\ \bibinfo {pages} {319} (\bibinfo {year} {2001})},\ \Eprint
  {http://arxiv.org/abs/astro-ph/0003442} {arXiv:astro-ph/0003442 [astro-ph]}
  \BibitemShut {NoStop}%
\bibitem [{\citenamefont {Greisen}(1966)}]{Greisen:1966jv}%
  \BibitemOpen
  \bibfield  {author} {\bibinfo {author} {\bibfnamefont {K.}~\bibnamefont
  {Greisen}},\ }\href {\doibase 10.1103/PhysRevLett.16.748} {\bibfield
  {journal} {\bibinfo  {journal} {Phys.Rev.Lett.}\ }\textbf {\bibinfo {volume}
  {16}},\ \bibinfo {pages} {748} (\bibinfo {year} {1966})}\BibitemShut
  {NoStop}%
\bibitem [{\citenamefont {Zatsepin}\ and\ \citenamefont
  {Kuzmin}(1966)}]{Zatsepin:1966jv}%
  \BibitemOpen
  \bibfield  {author} {\bibinfo {author} {\bibfnamefont {G.}~\bibnamefont
  {Zatsepin}}\ and\ \bibinfo {author} {\bibfnamefont {V.}~\bibnamefont
  {Kuzmin}},\ }\href@noop {} {\bibfield  {journal} {\bibinfo  {journal} {JETP
  Lett.}\ }\textbf {\bibinfo {volume} {4}},\ \bibinfo {pages} {78} (\bibinfo
  {year} {1966})}\BibitemShut {NoStop}%
\bibitem [{\citenamefont {Abu-Zayyad}\ \emph {et~al.}(2013)\citenamefont
  {Abu-Zayyad} \emph {et~al.}}]{Abu-Zayyad:2013dii}%
  \BibitemOpen
  \bibfield  {author} {\bibinfo {author} {\bibfnamefont {T.}~\bibnamefont
  {Abu-Zayyad}} \emph {et~al.} (\bibinfo {collaboration} {Telescope Array
  Collaboration}),\ }\href {\doibase 10.1103/PhysRevD.88.112005} {\bibfield
  {journal} {\bibinfo  {journal} {Phys.Rev.}\ }\textbf {\bibinfo {volume}
  {D88}},\ \bibinfo {pages} {112005} (\bibinfo {year} {2013})},\ \Eprint
  {http://arxiv.org/abs/1304.5614} {arXiv:1304.5614 [astro-ph.HE]} \BibitemShut
  {NoStop}%
\bibitem [{\citenamefont {Glushkov}\ \emph {et~al.}(2010)\citenamefont
  {Glushkov}, \citenamefont {Makarov}, \citenamefont {Pravdin}, \citenamefont
  {Sleptsov}, \citenamefont {Gorbunov}, \citenamefont {Rubtsov},\ and\
  \citenamefont {Troitsky}}]{Yak2010}%
  \BibitemOpen
  \bibfield  {author} {\bibinfo {author} {\bibfnamefont {A.}~\bibnamefont
  {Glushkov}}, \bibinfo {author} {\bibfnamefont {I.}~\bibnamefont {Makarov}},
  \bibinfo {author} {\bibfnamefont {M.}~\bibnamefont {Pravdin}}, \bibinfo
  {author} {\bibfnamefont {I.}~\bibnamefont {Sleptsov}}, \bibinfo {author}
  {\bibfnamefont {D.}~\bibnamefont {Gorbunov}}, \bibinfo {author}
  {\bibfnamefont {G.}~\bibnamefont {Rubtsov}}, \ and\ \bibinfo {author}
  {\bibfnamefont {S.}~\bibnamefont {Troitsky}} (\bibinfo {collaboration}
  {Yakutsk EAS Array}),\ }\href {\doibase 10.1103/PhysRevD.82.041101}
  {\bibfield  {journal} {\bibinfo  {journal} {Phys. Rev. D}\ }\textbf {\bibinfo
  {volume} {82}},\ \bibinfo {pages} {041101} (\bibinfo {year}
  {2010})}\BibitemShut {NoStop}%
\bibitem [{\citenamefont {Abraham}\ \emph {et~al.}(2009)\citenamefont {Abraham}
  \emph {et~al.}}]{Abraham:2009qb}%
  \BibitemOpen
  \bibfield  {author} {\bibinfo {author} {\bibfnamefont {J.}~\bibnamefont
  {Abraham}} \emph {et~al.} (\bibinfo {collaboration} {Pierre Auger
  Collaboration}),\ }\href {\doibase 10.1016/j.astropartphys.2009.04.003}
  {\bibfield  {journal} {\bibinfo  {journal} {Astropart.Phys.}\ }\textbf
  {\bibinfo {volume} {31}},\ \bibinfo {pages} {399} (\bibinfo {year} {2009})},\
  \Eprint {http://arxiv.org/abs/0903.1127} {arXiv:0903.1127 [astro-ph.HE]}
  \BibitemShut {NoStop}%
\bibitem [{\citenamefont {Settimo}()}]{Settimo}%
  \BibitemOpen
  \bibfield  {author} {\bibinfo {author} {\bibfnamefont {M.}~\bibnamefont
  {Settimo}} (\bibinfo {collaboration} {Pierre Auger}),\ }in\ \href@noop {}
  {\emph {\bibinfo {booktitle} {International Cosmic Ray
  Conference}}}\BibitemShut {NoStop}%
\bibitem [{\citenamefont {Aab}\ \emph {et~al.}(2014)\citenamefont {Aab} \emph
  {et~al.}}]{augercomp}%
  \BibitemOpen
  \bibfield  {author} {\bibinfo {author} {\bibfnamefont {A.}~\bibnamefont
  {Aab}} \emph {et~al.} (\bibinfo {collaboration} {Pierre Auger}),\ }\href
  {\doibase 10.1103/PhysRevD.90.122006} {\bibfield  {journal} {\bibinfo
  {journal} {Phys. Rev.}\ }\textbf {\bibinfo {volume} {D90}},\ \bibinfo {pages}
  {122006} (\bibinfo {year} {2014})},\ \Eprint {http://arxiv.org/abs/1409.5083}
  {arXiv:1409.5083 [astro-ph.HE]} \BibitemShut {NoStop}%
\bibitem [{\citenamefont {Siohan}\ \emph {et~al.}(1978)\citenamefont {Siohan},
  \citenamefont {Ellsworth}, \citenamefont {Ito}, \citenamefont {Macfall},
  \citenamefont {Streitmatter} \emph {et~al.}}]{Siohan:1978zk}%
  \BibitemOpen
  \bibfield  {author} {\bibinfo {author} {\bibfnamefont {F.}~\bibnamefont
  {Siohan}}, \bibinfo {author} {\bibfnamefont {R.}~\bibnamefont {Ellsworth}},
  \bibinfo {author} {\bibfnamefont {A.}~\bibnamefont {Ito}}, \bibinfo {author}
  {\bibfnamefont {J.}~\bibnamefont {Macfall}}, \bibinfo {author} {\bibfnamefont
  {R.}~\bibnamefont {Streitmatter}},  \emph {et~al.},\ }\href {\doibase
  10.1088/0305-4616/4/7/021} {\bibfield  {journal} {\bibinfo  {journal}
  {J.Phys.}\ }\textbf {\bibinfo {volume} {G4}},\ \bibinfo {pages} {1169}
  (\bibinfo {year} {1978})}\BibitemShut {NoStop}%
\bibitem [{\citenamefont {Honda}\ \emph {et~al.}(1993)\citenamefont {Honda},
  \citenamefont {Nagano}, \citenamefont {Tonwar}, \citenamefont {Kasahara},
  \citenamefont {Hara} \emph {et~al.}}]{Honda1992}%
  \BibitemOpen
  \bibfield  {author} {\bibinfo {author} {\bibfnamefont {M.}~\bibnamefont
  {Honda}}, \bibinfo {author} {\bibfnamefont {M.}~\bibnamefont {Nagano}},
  \bibinfo {author} {\bibfnamefont {S.}~\bibnamefont {Tonwar}}, \bibinfo
  {author} {\bibfnamefont {K.}~\bibnamefont {Kasahara}}, \bibinfo {author}
  {\bibfnamefont {T.}~\bibnamefont {Hara}},  \emph {et~al.},\ }\href {\doibase
  10.1103/PhysRevLett.70.525} {\bibfield  {journal} {\bibinfo  {journal}
  {Phys.Rev.Lett.}\ }\textbf {\bibinfo {volume} {70}},\ \bibinfo {pages} {525}
  (\bibinfo {year} {1993})}\BibitemShut {NoStop}%
\bibitem [{\citenamefont {Knurenkoa}\ \emph {et~al.}(2013)\citenamefont
  {Knurenkoa}, \citenamefont {Sabourovb} \emph {et~al.}}]{Knurenko2013}%
  \BibitemOpen
  \bibfield  {author} {\bibinfo {author} {\bibfnamefont {S.}~\bibnamefont
  {Knurenkoa}}, \bibinfo {author} {\bibfnamefont {A.}~\bibnamefont
  {Sabourovb}},  \emph {et~al.},\ }\href {\doibase 10.1051/epjconf/20135307006}
  {\bibfield  {journal} {\bibinfo  {journal} {EPJ Web of Conferences}\ }\textbf
  {\bibinfo {volume} {53}},\ \bibinfo {pages} {07006} (\bibinfo {year}
  {2013})}\BibitemShut {NoStop}%
\bibitem [{\citenamefont {Mielke}\ \emph {et~al.}(1994)\citenamefont {Mielke},
  \citenamefont {Foeller}, \citenamefont {Engler},\ and\ \citenamefont
  {Knapp}}]{Mielke:1994un}%
  \BibitemOpen
  \bibfield  {author} {\bibinfo {author} {\bibfnamefont {H.}~\bibnamefont
  {Mielke}}, \bibinfo {author} {\bibfnamefont {M.}~\bibnamefont {Foeller}},
  \bibinfo {author} {\bibfnamefont {J.}~\bibnamefont {Engler}}, \ and\ \bibinfo
  {author} {\bibfnamefont {J.}~\bibnamefont {Knapp}},\ }\href {\doibase
  10.1088/0954-3899/20/4/010} {\bibfield  {journal} {\bibinfo  {journal}
  {J.Phys.}\ }\textbf {\bibinfo {volume} {G20}},\ \bibinfo {pages} {637}
  (\bibinfo {year} {1994})}\BibitemShut {NoStop}%
\bibitem [{\citenamefont {Belov}(2006)}]{Belov2006}%
  \BibitemOpen
  \bibfield  {author} {\bibinfo {author} {\bibfnamefont {K.}~\bibnamefont
  {Belov}} (\bibinfo {collaboration} {HiRes Collaboration}),\ }\href {\doibase
  10.1016/j.nuclphysbps.2005.07.035} {\bibfield  {journal} {\bibinfo  {journal}
  {Nucl.Phys.Proc.Suppl.}\ }\textbf {\bibinfo {volume} {151}},\ \bibinfo
  {pages} {197} (\bibinfo {year} {2006})}\BibitemShut {NoStop}%
\bibitem [{\citenamefont {Aielli}\ \emph {et~al.}(2009)\citenamefont {Aielli}
  \emph {et~al.}}]{Aielli:2009ca}%
  \BibitemOpen
  \bibfield  {author} {\bibinfo {author} {\bibfnamefont {G.}~\bibnamefont
  {Aielli}} \emph {et~al.} (\bibinfo {collaboration} {ARGO-YBJ
  Collaboration}),\ }\href {\doibase 10.1103/PhysRevD.80.092004} {\bibfield
  {journal} {\bibinfo  {journal} {Phys.Rev.}\ }\textbf {\bibinfo {volume}
  {D80}},\ \bibinfo {pages} {092004} (\bibinfo {year} {2009})},\ \Eprint
  {http://arxiv.org/abs/0904.4198} {arXiv:0904.4198 [hep-ex]} \BibitemShut
  {NoStop}%
\bibitem [{\citenamefont {Aglietta}\ \emph {et~al.}(2009)\citenamefont
  {Aglietta} \emph {et~al.}}]{EAStop}%
  \BibitemOpen
  \bibfield  {author} {\bibinfo {author} {\bibfnamefont {M.}~\bibnamefont
  {Aglietta}} \emph {et~al.} (\bibinfo {collaboration} {EAS-TOP
  Collaboration}),\ }\href {\doibase 10.1103/PhysRevD.79.032004} {\bibfield
  {journal} {\bibinfo  {journal} {Phys. Rev. D}\ }\textbf {\bibinfo {volume}
  {79}},\ \bibinfo {pages} {032004} (\bibinfo {year} {2009})}\BibitemShut
  {NoStop}%
\bibitem [{\citenamefont {Engel}\ \emph {et~al.}(1998)\citenamefont {Engel},
  \citenamefont {Gaisser}, \citenamefont {Lipari},\ and\ \citenamefont
  {Stanev}}]{Engel:1998pw}%
  \BibitemOpen
  \bibfield  {author} {\bibinfo {author} {\bibfnamefont {R.}~\bibnamefont
  {Engel}}, \bibinfo {author} {\bibfnamefont {T.}~\bibnamefont {Gaisser}},
  \bibinfo {author} {\bibfnamefont {P.}~\bibnamefont {Lipari}}, \ and\ \bibinfo
  {author} {\bibfnamefont {T.}~\bibnamefont {Stanev}},\ }\href {\doibase
  10.1103/PhysRevD.58.014019} {\bibfield  {journal} {\bibinfo  {journal}
  {Phys.Rev.}\ }\textbf {\bibinfo {volume} {D58}},\ \bibinfo {pages} {014019}
  (\bibinfo {year} {1998})},\ \Eprint {http://arxiv.org/abs/hep-ph/9802384}
  {arXiv:hep-ph/9802384 [hep-ph]} \BibitemShut {NoStop}%
\bibitem [{\citenamefont {Gaisser}\ \emph {et~al.}(1987)\citenamefont
  {Gaisser}, \citenamefont {Sukhatme},\ and\ \citenamefont
  {Yodh}}]{Gaisser:1986haa}%
  \BibitemOpen
  \bibfield  {author} {\bibinfo {author} {\bibfnamefont {T.}~\bibnamefont
  {Gaisser}}, \bibinfo {author} {\bibfnamefont {U.}~\bibnamefont {Sukhatme}}, \
  and\ \bibinfo {author} {\bibfnamefont {G.}~\bibnamefont {Yodh}},\ }\href
  {\doibase 10.1103/PhysRevD.36.1350} {\bibfield  {journal} {\bibinfo
  {journal} {Phys.Rev.}\ }\textbf {\bibinfo {volume} {D36}},\ \bibinfo {pages}
  {1350} (\bibinfo {year} {1987})}\BibitemShut {NoStop}%
\bibitem [{\citenamefont {Glauber}\ and\ \citenamefont
  {Matthiae}(1970)}]{Glauber70}%
  \BibitemOpen
  \bibfield  {author} {\bibinfo {author} {\bibfnamefont {R.}~\bibnamefont
  {Glauber}}\ and\ \bibinfo {author} {\bibfnamefont {G.}~\bibnamefont
  {Matthiae}},\ }\href@noop {} {\bibfield  {journal} {\bibinfo  {journal}
  {Nucl.Phys.}\ }\textbf {\bibinfo {volume} {B21}},\ \bibinfo {pages} {135}
  (\bibinfo {year} {1970})}\BibitemShut {NoStop}%
\bibitem [{\citenamefont {Dias~de Deus}\ and\ \citenamefont
  {Kroll}(1978)}]{DiasdeDeus1978}%
  \BibitemOpen
  \bibfield  {author} {\bibinfo {author} {\bibfnamefont {J.}~\bibnamefont
  {Dias~de Deus}}\ and\ \bibinfo {author} {\bibfnamefont {P.}~\bibnamefont
  {Kroll}},\ }\href@noop {} {\bibfield  {journal} {\bibinfo  {journal} {Acta
  Phys. Polon.}\ }\textbf {\bibinfo {volume} {B9}},\ \bibinfo {pages} {157}
  (\bibinfo {year} {1978})}\BibitemShut {NoStop}%
\bibitem [{\citenamefont {Buras}\ and\ \citenamefont {Dias~de
  Deus}(1974)}]{Buras:1973km}%
  \BibitemOpen
  \bibfield  {author} {\bibinfo {author} {\bibfnamefont {A.}~\bibnamefont
  {Buras}}\ and\ \bibinfo {author} {\bibfnamefont {J.}~\bibnamefont {Dias~de
  Deus}},\ }\href {\doibase 10.1016/0550-3213(74)90197-7} {\bibfield  {journal}
  {\bibinfo  {journal} {Nucl.Phys.}\ }\textbf {\bibinfo {volume} {B71}},\
  \bibinfo {pages} {481} (\bibinfo {year} {1974})}\BibitemShut {NoStop}%
\bibitem [{\citenamefont {Donnachie}\ and\ \citenamefont
  {Landshoff}(1992)}]{Donnachie:1992ny}%
  \BibitemOpen
  \bibfield  {author} {\bibinfo {author} {\bibfnamefont {A.}~\bibnamefont
  {Donnachie}}\ and\ \bibinfo {author} {\bibfnamefont {P.}~\bibnamefont
  {Landshoff}},\ }\href {\doibase 10.1016/0370-2693(92)90832-O} {\bibfield
  {journal} {\bibinfo  {journal} {Phys.Lett.}\ }\textbf {\bibinfo {volume}
  {B296}},\ \bibinfo {pages} {227} (\bibinfo {year} {1992})},\ \Eprint
  {http://arxiv.org/abs/hep-ph/9209205} {arXiv:hep-ph/9209205 [hep-ph]}
  \BibitemShut {NoStop}%
\bibitem [{\citenamefont {Avila}\ \emph {et~al.}(1999)\citenamefont {Avila}
  \emph {et~al.}}]{pp1999}%
  \BibitemOpen
  \bibfield  {author} {\bibinfo {author} {\bibfnamefont {C.}~\bibnamefont
  {Avila}} \emph {et~al.},\ }\href {\doibase 10.1016/S0370-2693(98)01421-X}
  {\bibfield  {journal} {\bibinfo  {journal} {Phys.Lett. B}\ }\textbf {\bibinfo
  {volume} {445}},\ \bibinfo {pages} {419–422} (\bibinfo {year}
  {1999})}\BibitemShut {NoStop}%
\bibitem [{\citenamefont {Antchev}\ \emph {et~al.}(2011)\citenamefont
  {Antchev}, \citenamefont {Aspell}, \citenamefont {Atanassov}, \citenamefont
  {Avati}, \citenamefont {Baechler} \emph {et~al.}}]{Totem2011}%
  \BibitemOpen
  \bibfield  {author} {\bibinfo {author} {\bibfnamefont {G.}~\bibnamefont
  {Antchev}}, \bibinfo {author} {\bibfnamefont {P.}~\bibnamefont {Aspell}},
  \bibinfo {author} {\bibfnamefont {I.}~\bibnamefont {Atanassov}}, \bibinfo
  {author} {\bibfnamefont {V.}~\bibnamefont {Avati}}, \bibinfo {author}
  {\bibfnamefont {J.}~\bibnamefont {Baechler}},  \emph {et~al.},\ }\href
  {\doibase 10.1209/0295-5075/96/21002} {\bibfield  {journal} {\bibinfo
  {journal} {Europhys.Lett.}\ }\textbf {\bibinfo {volume} {96}},\ \bibinfo
  {pages} {21002} (\bibinfo {year} {2011})},\ \Eprint
  {http://arxiv.org/abs/1110.1395} {arXiv:1110.1395 [hep-ex]} \BibitemShut
  {NoStop}%
\end{thebibliography}
%

\end{document}